*Title: Insular Microbiogeography*

Running title: Insular microbiogeography

Authors: James H. Kaufman[1]*, Christopher A. Elkins[2], Matthew Davis[1], Allison M Weis[3], Bihua C. Huang [3], Mark K Mammel[2], Isha R. Patel[2], Kristen L. Beck[1], Stefan Edlund[1], David Chambliss[1], Simone Bianco[1], Mark Kunitomi[1], Bart C. Weimer[3*]

[1]IBM Almaden Research Center, 650 Harry Rd., San Jose, CA, USA; [2]Division of Molecular Biology, Center for Food Safety and Applied Nutrition, United States Food and Drug Administration, Laurel, MD 20708 USA; [3]University of California, Davis, School of Veterinary Medicine, 100K Pathogen Genome Project, 1089 Veterinary Medicine Dr., Davis, CA 95616 USA

| | |
|---|---|
| James H. Kaufman | <jhkauf@us.ibm.com> |
| Christopher A. Elkins | <chris.elkins@fda.hhs.gov> |
| Matthew Davis | <mattadav@us.ibm.com> |
| Allison M. Weis | <amweiss@ucdavis.edu> |
| Bihua C. Huang | <bcahuang@ucdavis.edu> |
| Mark K. Mammel | <Mark.Mammel@fda.hhs.gov> |
| Isha Patel | <Isha.Patel@fda.hhs.gov> |
| Kristen L. Beck | <klbeck@us.ibm.com> |
| Stefan Edlund | <sedlund@us.ibm.com> |
| David Chambliss | <chamb@us.ibm.com> |
| Simone Bianco | <sbianco@us.ibm.com> |
| Mark Kunitomi | <Mark.Kunitomi@ibm.com> |
| Bart C. Weimer | <bcweimer@ucdavis.edu> |

**Co-corresponding authors:* James Kaufman, jhkauf@us.ibm.com; Bart C. Weimer, bcweimer@ucdavis.edu

**Contributions:** JHK conceived the theory, designed and conducted experiments, and wrote manuscript; CAE conducted laboratory experiments, contributed to manuscript; MD conducted experiments, edited manuscript; AMW conducted experiments, edited manuscript; BCH conducted laboratory experiments, edited manuscript; MKM conducted experiments, edited manuscript; IP conducted experiments, edited manuscript; KLB conducted experiments, edited manuscript; SE conducted experiments, edited manuscript; DC conducted experiments, edited manuscript; SB conducted experiments, contributed to manuscript; BCW contributed to the theory, designed and conducted experiments, and wrote manuscript.

# I. ABSTRACT

The diversity revealed by large scale genomics in microbiology is calling into question long held beliefs about genome stability, evolutionary rate, even the definition of a species. MacArthur and Wilson's theory of insular biogeography provides an explanation for the diversity of macroscopic animal and plant species as a consequence of the associated hierarchical web of species interdependence. We report a large scale study of microbial diversity that reveals that the cumulative number of genes discovered increases with the number of genomes studied as a simple power law. This result is demonstrated for three different genera comparing over 15,000 isolates. We show that this power law is formally related to the MacArthur-Wilson exponent, suggesting the emerging diversity of microbial genotypes arises because the scale independent behavior first reported by MacArthur and Wilson extends down to the scale of microbes and their genes. Assessing the depth of available whole genome sequences implies a dynamically changing core genome, suggesting that traditional taxonomic classifications should be replaced with a quasispecies model that captures the diversity and dynamic exchange of genes. We report Species population "clouds" in a defined microbiome, with scale invariance extending down to the level of single-nucleotide polymorphisms (SNPs).

# II. INTRODUCTION

Microbiologists are now engaged in describing the diversity of microbial life on earth utilizing technological advances in high throughput sequencing. A growing body of Whole Genome Sequencing (WGS) data is accessible for scientific research in public repositories, including the National Center for Biotechnology Information (NCBI), Sequence Read Archive (SRA), the European Nucleotide Archive (ENA), the DNA Data Bank of Japan (DDBJ), the Genomic Encyclopedia of Bacteria and Archaea (GEBA), and the multinational 100K Pathogen Genome Project [1-3]. The crowd-sourced data in these WGS databases has resulted in cataloguing of a plethora of microbial genomes and has assisted in the examination of the metagenomics of ecological niches important in medicine, agriculture, energy, and the built environment [4-6].

This implementation of large scale genomics in microbiology gives rise to observations previously impossible. Increased genome diversity challenges long held beliefs about genome stability, evolutionary rate, and the definition of a species [7,8]. Indeed, the number of unique genes that are represented by a species, known as the pan-genome, continues to increase as the number of genes conserved across all the members of a species, known as the core-genome, diminishes. The forces that define this trajectory of novel gene discovery evoke consideration of the advances made by naturalists when examining diversity of macroscopic life.

MacArthur and Wilson's macroecological "Theory of Island Biogeography" explains species richness and diversity across geographically disconnected groups [9] [10]. This theory relates the number of species ($S$) found in steady state within the area ($A$) of an isolated ecosystem as a power law, where $S$ varies as $A^z$ with $z$, the MacArthur-Wilson exponent, typically in the range of $0.2<z<0.3$, slightly decreasing for island groups nearer to continental land masses.

This power law, observed over five orders of area magnitude, defines a range over which the species-area relationship is mathematically invariant, i.e., the exponent does not change. This is commonly referred to as "scale-free" behavior. The relationship is also robust against perturbation. For example, species diversity collapses exponentially after clear-cutting Amazon forests, but the steady state species diversity remaining in isolated reserves obeys the MacArthur-Wilson theory of insular biogeography with the same exponent, $z$ [11].

Evidence for scale-free behavior is observed across many fields of science and many orders of scale, ranging from microscopic study of porous materials to large scale studies in geology and seismology [12]. While infinite self-similarity is only a mathematical ideal, experimental evidence for scale invariance over a wide range of scale in natural systems often gives insights into underlying processes that govern interactions within those systems. For example, fractal geometry and fractional exponents do not instantaneously appear in the world, but rather evolve based on underlying dynamic processes [13]. Ecological models based on the work of Wilson and MacArthur describe how the steady state species-area relationship emerges because the rate of successful immigration of new species (from distant continental reservoirs) and the rate of species extinction both depend upon the availability of viable niches. The number of niches depend, in turn, on the available land area. Evolutionary models support this theory and show, as predicted by Wilson, that development of the food web leads to the observed scaling relationship as successful species provide additional niches for new immigrants [14]. In these models, the scale-free behavior emerges from the associated hierarchical web of species interdependence (i.e., the food web or microbial community structure) [13,14].

The diversity of microbes implied by MacArthur and Wilson's theory and empirically observed through the sequencing of hundreds of thousands of genomes requires us to reframe the system of classification used to define species. Eigen and Schuster suggest that a "cloud" of diverse but related organisms within a population are more accurately represented as "quasispecies" with a distribution of genomes and a distribution of genes [15,16]. This concept is commonly invoked in virology but data demonstrating its applications to bacterial evolution has been limited [2,17]. Available genome sequences are just now reaching the critical mass required to make the observations demonstrated here [1,2,18,19].

In addition to mutation, recombination and horizontal gene transfer are known to play an important role in genetic diversity bacteria [7]. Recent studies have shown some bacterial populations exhibit exceptionally high intrinsic mutation and recombination rates, either throughout the course of infection [20] or during the acute phase of infection [21]. Increased mutation rates in bacteria are also apparent during selection pressure, either by natural predators [22], or through expression of alternative error-prone mutator genes [23]. Several *Campylobacter* strains are known to show increased mutation rates [24], with instances of *C. jejuni* and *C. coli* hypermutator phenotypes linked to the emergence of ciproxin resistance.

Here, we report a large scale study of genomic diversity of genes for thousands of isolates of three different bacterial genera. Our work reveals scaling of the number of new genes vs. the number of samples or genotypes. We also derive a "scaling law" that relates the exponent measured here to the MacArthur and Wilson exponent relating species or taxa to *geographic area*. Based on this relationship we are able to accurately measure the MacArthur-Wilson exponent, *z*. Our analyses reveal a common exponent for closely related bacteria (*Salmonella* and *E. coli*) and a slightly larger exponent for a distant organism (*Campylobacter*) with higher mutation rate and a smaller genome. All the exponents are in agreement with the range observed by Wilson and MacArthur. Although previous studies have failed to find evidence for scale-free behavior in bacterial diversity, they may have examined environmental samples from areas that were not truly insular.[25-27]

# III. RESULTS

## A. Diversity of genes

To quantify diversity in terms of gene content, we did large scale comparison of individual proteins from genome datasets of three different organisms, *Salmonella*, *E. coli*, and *Campylobacter*. Samples were derived from publically available deep sequencing datasets where each isolate was cultured, isolated, and sequenced as clonal colonies. We performed a de novo assembly and gene annotation for each isolate and then quantified the number of total genes as a function of number of isolates sampled in multiple bootstrapped trials. Each point in Figure 1 represents the cumulative number of distinct amino acid sequences inferred for regions annotated as known or hypothetical proteins for one genome subset, plotted against the number of isolates in that subset, both known and unknown. The points are derived from separate bootstrapped trials (see Appendix A: Methods).

For *Salmonella,* log of the genome diversity grows approximately linearly with the log of the number of isolates in Figure 1a, indicating a power law relationship with the average rate of cumulative gene discovery *(N*) increasing with the number of isolates, $\eta$*,* as:

$$N \propto \eta^{0.468\pm0.001} \qquad \text{Eq. 1 (\textit{Salmonella})}$$

The same power law (within experimental error) is observed in an analysis of the *E. coli* pangenome (Figure 1b). In this study, 3348 independent genome sequences were first classified into *334 genotypes* to condense the data and search for a common reference representative with < 0.2% genetic difference. Conducted to establish a "core" gene set common to all 3348 genomes, this analysis was different from the core analysis used to construct a phylogeny of the species (which consisted of 42 closed genomes; see Methods). As with *Salmonella* (Figure 1a), a power law emerges for *E. coli* relating the average rate of cumulative gene discovery, $N$, to the number of genotypes, $\eta$, as:

$$N \propto \eta^{0.462 \pm 0.002} \qquad \text{Eq. 2 } (E.\ coli)$$

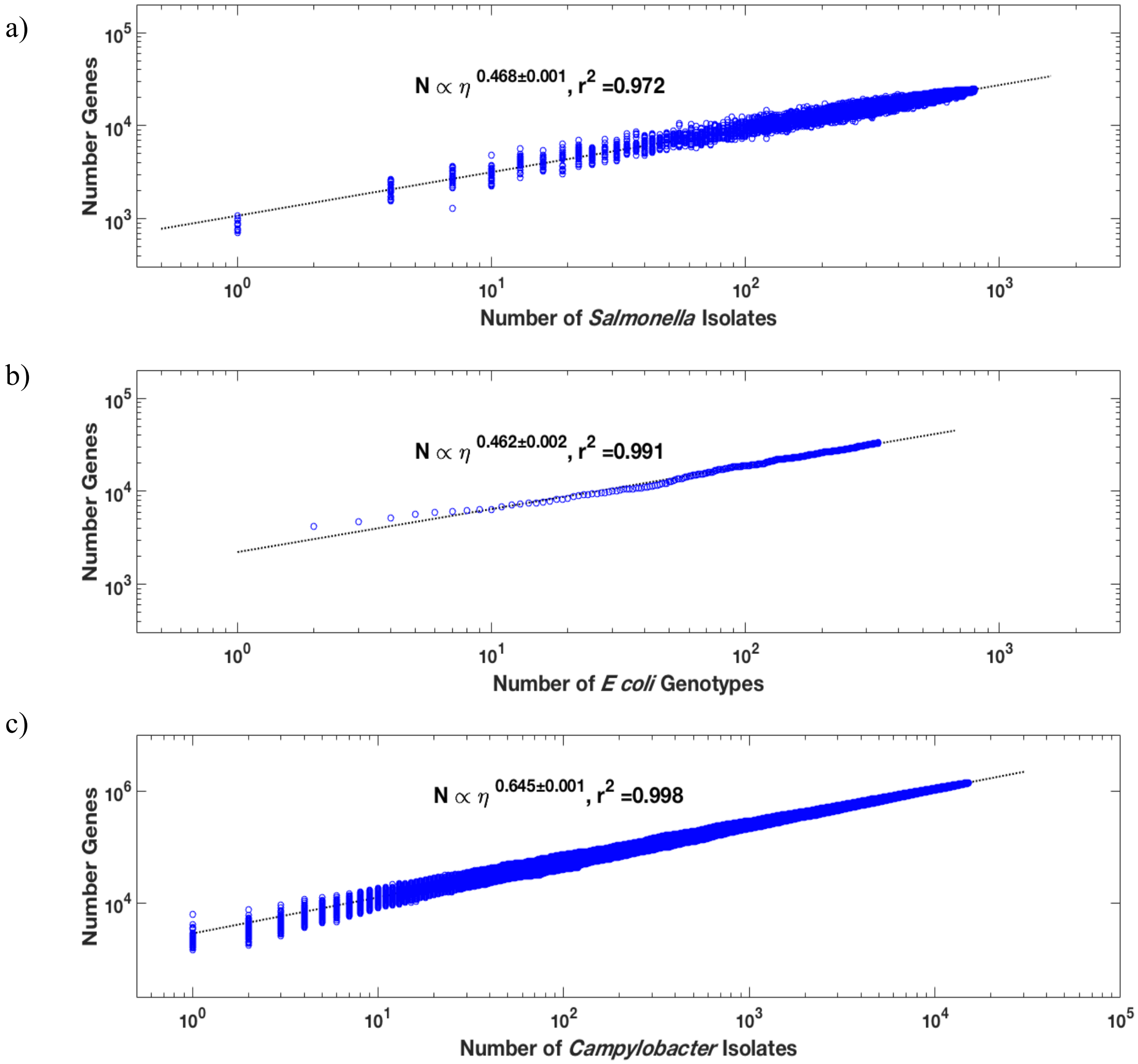

**Figure 1.** The cumulative rate of discovery of "new genes" as a function of the number of *Salmonella* isolates (a), *E. coli* genotypes (b), and *Campylobacter* isolates (c) *all* increase as a power law (linear on a log-log plot).

To determine whether genome size and/or mutation rate contributes to variation in the exponent, we extended this investigation to another organism, *Campylobacter* (Figure 1c). *Campylobacter* is often found in the same niche (e.g., chicken microbiome) and known to have a much higher mutation rate as compared to *E. coli* and *Salmonella* (which have similar mutation rates). To compensate for the additional genetic inclusion rates we also increased the number of genomes to 15,158. This greatly expands scale of the genomic calculation. The data (Figure 1c) once again follows an approximate power law relating the rate of cumulative gene discovery, N, to the number of genotypes, $\eta$, as:

$$N \propto \eta^{0.645 \pm 0.001} \qquad \text{Eq. 3 } (\textit{Campylobacter})$$

In Figure 1, we deliberately use the same symbol ($\eta$) to denote both number of isolates (Figures 1 a,c) and number of genotypes (Figure 1b). Each genotype representing a collection of isolates grouped in a data reduction effort. The power law behavior observed in these studies of three completely different genera, using two independent methods of analysis, suggests a possibly universal functional relationship between the rate of new gene discovery and the number of genotypes or isolates over a wide range in scale.

## B. Taxonomic context

As the number of bacterial WGS increases to over 330,000, the emerging genotypic diversity is challenging taxonomic designations and concepts of evolution [28]. Classification and naming of organisms with traditional methods are becoming problematic, as addition of a single new genome drives continual revision of reference phylogenetic trees. For example, recently discovered cryptic environmental lineages of *E. coli* do not fit with current multi locus sequence typing (MLST) classification for *E. coli sensu stricto* [29]. WGS analysis with a single gene, such as 16s rRNA, provides one name, but whole genome alignment provides a mixture of names where portions of the genome align to >1 genome, demonstrating genome plasticity or widespread homology among gene groups. A similar conclusion was hinted at in a study of lactic acid bacteria (LAB) where two entire genera were re-aligned based on the largest LAB comparison of the time [30]. As the volume of WGS data continues to grow, some conventional methods will not scale to meet the standard of absolute identification. Whole genome comparison methods now available circumvent the assumptions needed to define a constant or static set of single genes, SNPs, or even core sets of genes; these methods eliminate the need for data reduction and avoid the associated skewing of taxonomical classification. Although there is a regulatory need to classify and name organisms, the data in this paper suggests that classification of bacteria, based on whole genomes, should be considered more akin to the classification of virus as a quasispecies, as discussed below.

To illustrate how whole genome data challenges traditional taxonomy, we provide an analysis of three independent *Campylobacter* data sets at different data scales. Principal component analysis[31] based on a whole genome-genome difference matrix reveals similar clusters of *Campylobacter* genomes as shown in Figures 2a-c [3,32]. When using only 90 genomes from a study of primates and crows (Figure 2a), three tight clusters appear in the PCA biplot (Figure 2a); this result aligns well with classic phylogenetic classification and nomenclature, and is congruent with a small sampling of the genome space based on only a few isolates. Using 218 genomes from the Ensembl database (Figure 2b), the clusters for the three species identified in Figure 2a expand (and overlap) substantially, demonstrating the increased genome diversity for an increase of approximately two times more isolates in the analysis. Using 715 genomes from the NCBI SRA (Figure 2c) in the same analysis, the *Campylobacter* genome diversity becomes even more apparent. Taken together, these data demonstrate a rapid expansion of genotypic diversity within species, and of diversification, especially in *C. jejuni*, that overlaps other species in genome space. The whole genome analysis reveals how a relatively small increase in the number of genomes can rapidly expand genome space and blur the concept of species.

a)

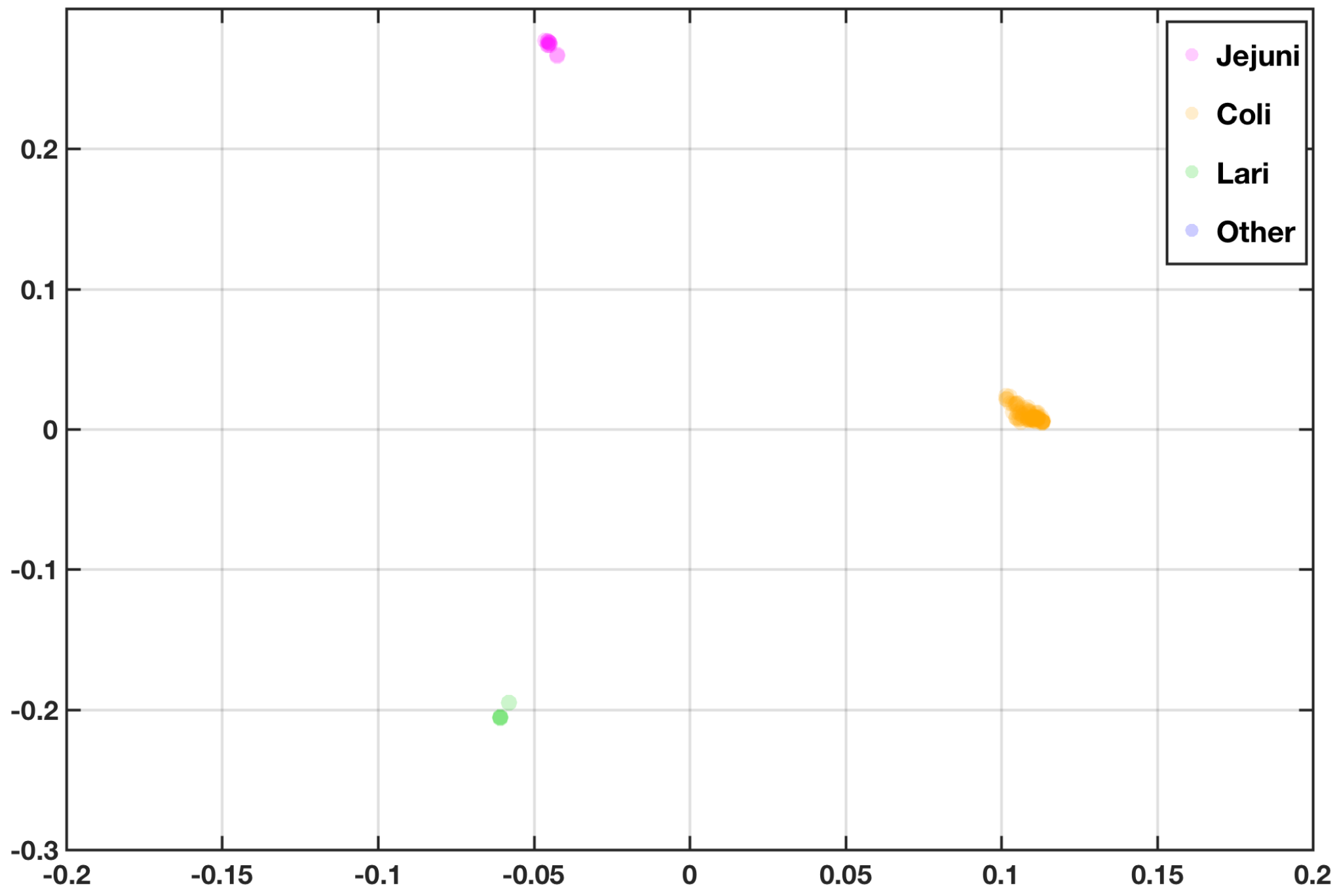

b)

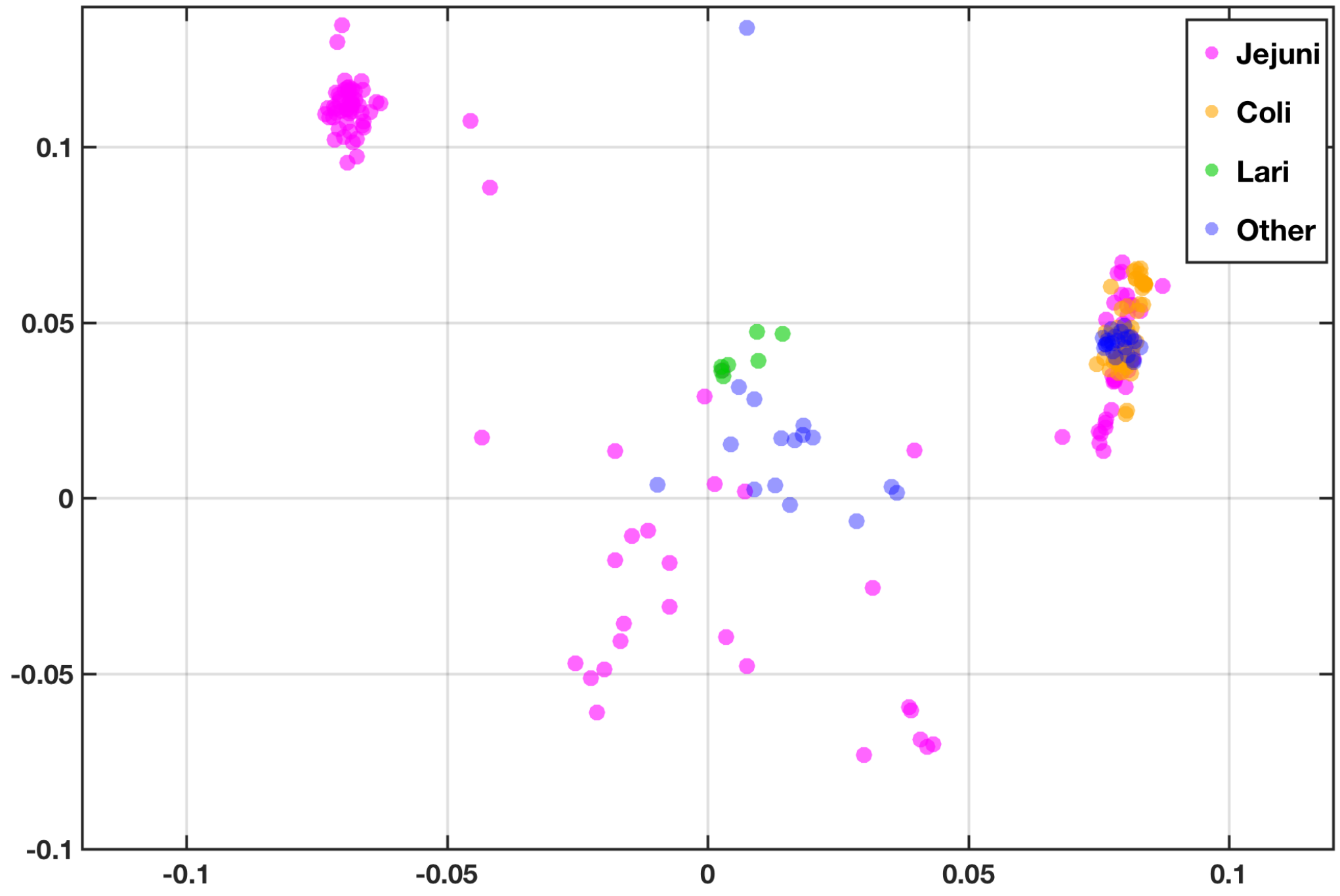

c)

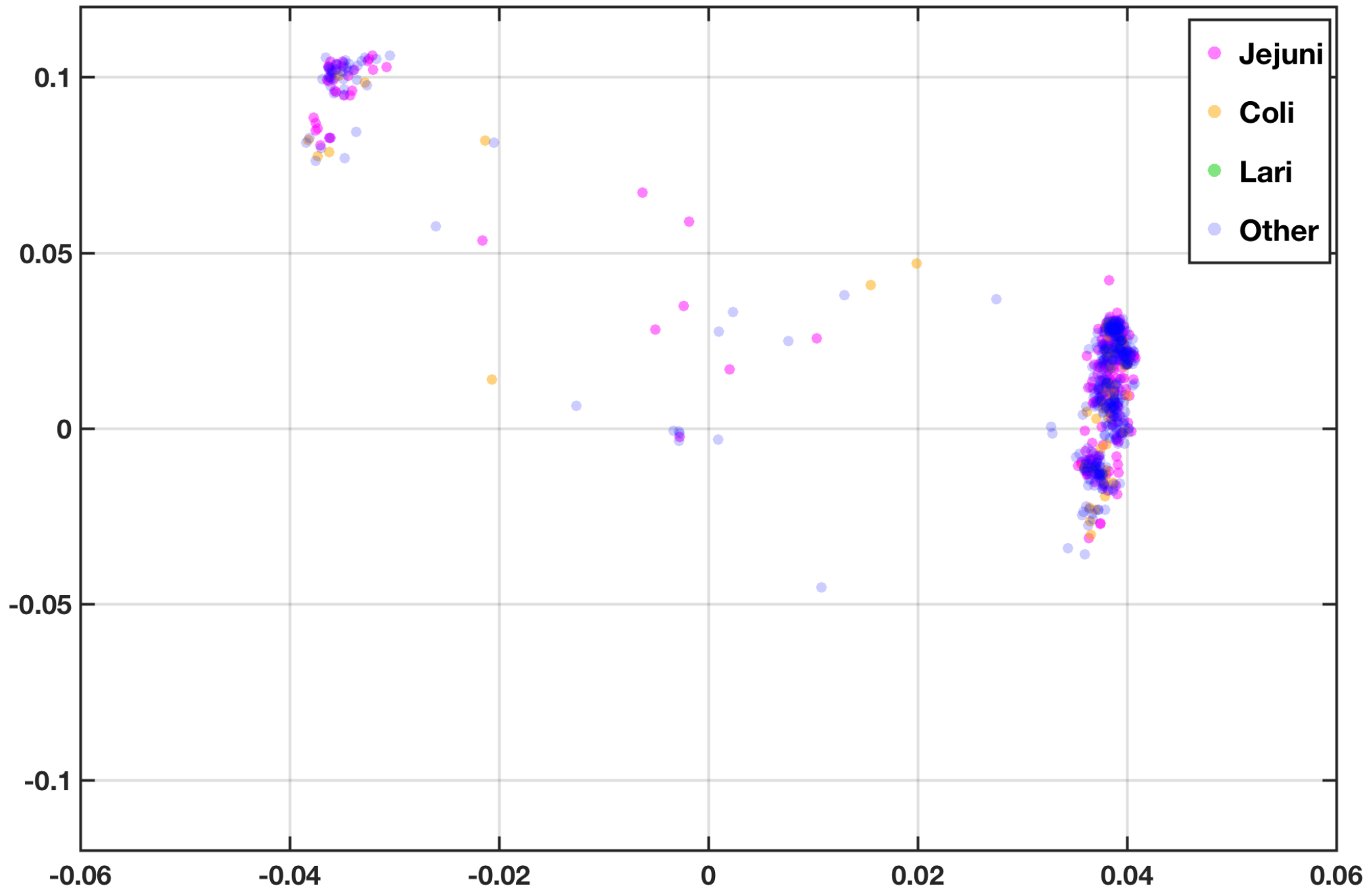

**Figure 2a-c:** Principal component analyses of genome-genome distance from *Campylobacter* genomes representing (a) 90 genomes from a primates and crows [32], (b) 218 genomes from the Ensembl reference database, and (c) 715 genomes from the NCBI Sequence Read Archive (SRA). In all three figures the x axis is principle component 1, and the y axis principle component 2. The color code in the legend indicates the original source species identification. With increasing dataset size, the clades representing traditional views of phylogeny for species begin to expand and overlap.

This expansion in diversity reflects the increasing genome dis-similarity that occurs as more genomes are added to an analysis. Expansion of the clades is especially evident when samples originate from a large collection of organisms from different geographical locations that may be insular regions of diversity [9,32]. The expansion and overlap in Figure 2 implies that traditional naming conventions are inaccurate and lead to false identification based on artifacts that emerge when using a small set of input genes, MLST markers, or single nucleotide polymorphisms (SNPs).

We also observe intersecting taxonomic grouping in the full genome-genome similarity matrix, rendered as a heat map in Figures 3a-c. We measured all pairwise distances between *Campylobacter* whole genomes. With 90 genomes (Figure 3a) the regions of greatest similarity are found along the matrix diagonal. With 218 and 715 genomes (Figures 3b and 3c respectively), highly similar but divergent sub-sets of genomes emerge. These appear as regions of similarity for well separated (off-diagonal) genome pairs, indicating similarity between divergent branches of the taxonomic graph. Fine structure emerges within the clades as more genomes are added to the analysis. When 715 genomes were used (Figure 3c), the matrix reveals regions of genetic similarity even for more distant taxa, indicating gene transfer between different species. Analogous behavior is observed in divergent branches of *E. coli* and *Salmonella enterica* (See Appendix B, Figure B1). Shapiro attributed this diversity expansion to horizontal gene transfer [5], while Doolittle prophetically described gene sharing beyond what was thought possible [7]. Subsequently, many studies have established that bacteria engage in extensive horizontal gene transfer [7,8,33,34]. Horizontal gene transfer suggests that "the tree of life" is not a simple tree [34]. The diversity apparent in Figures 1-3 suggest that bacterial life is organized in a highly interconnected network, a graph containing edges that connect phylogenetic quasispecies across varying genetic distances or scales.

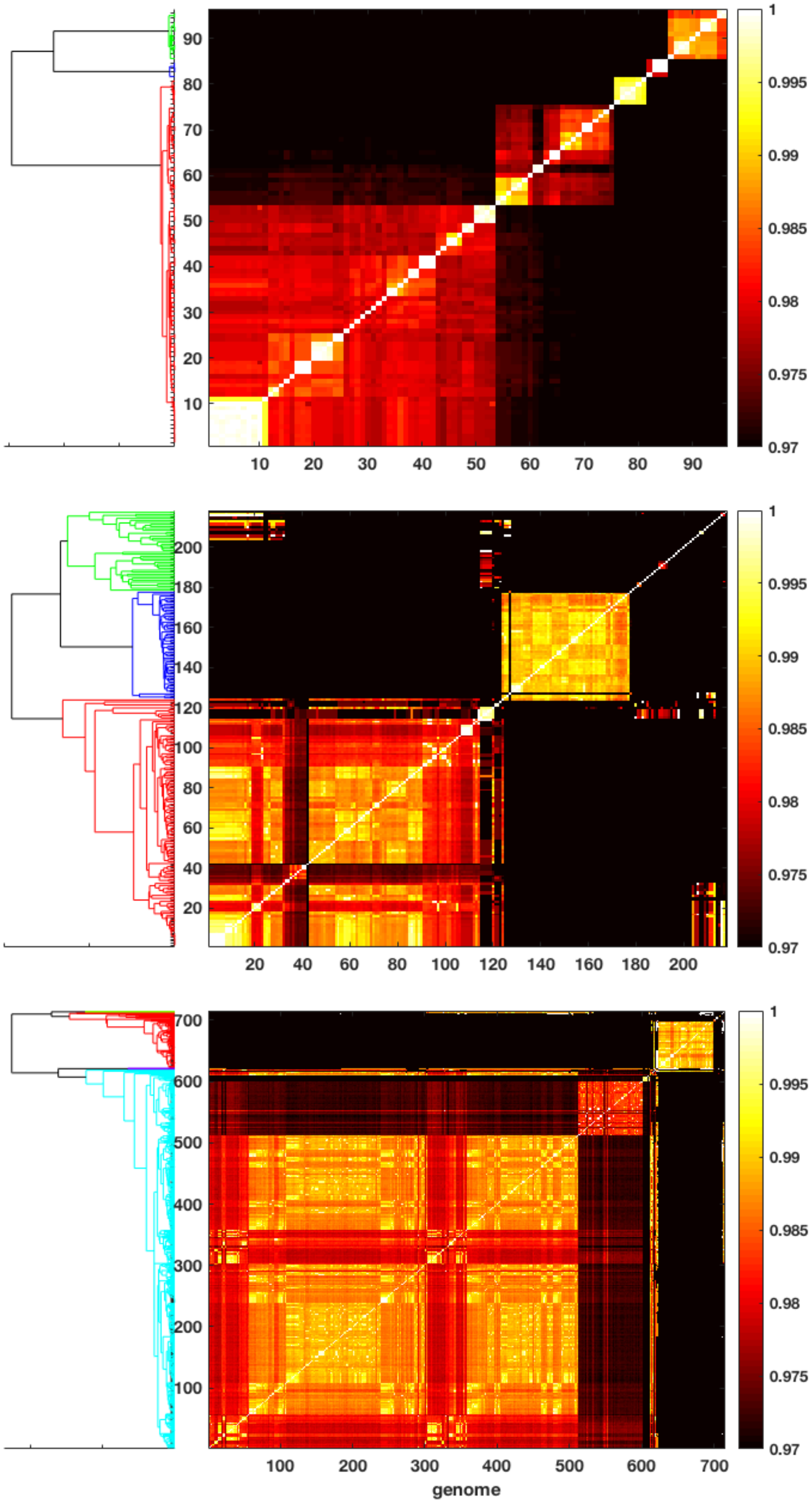

**Figure 3a-c:** Heat maps showing the genome-genome similarity matrix derived from three *Campylobacter* databases containing (a) 90 genomes, (b) 218 genomes, and (c) 715 genomes.

## C. The quasispecies model

Eigen and Schuster suggest that the "cloud" of diverse but related organisms within a population are more accurately represented as "quasispecies" with a distribution of genomes and a distribution of genes [15,16]. This concept is commonly invoked in virology but data demonstrating its applications to bacterial evolution has been limited [2,17]. Available genome sequences are just now reaching a critical mass required to make the observations demonstrated here [1,2,18,19].

In the experiments described above, samples were derived using laboratory procedures where each pathogen is studied as a single genotype, reflecting a traditional view of phylogeny where organisms are cultured, isolated, and sequenced as clonal colonies. In fact, organisms exist in genomically diverse populations or communities of related, but not identical genotypes – even within a colony. These communities evolve (rapidly when under pressure) by a number of mechanisms including horizontal gene transfer, methylation, and plasmid content. Culture organisms for sequencing as reference genomes are recognized as the *average genome of a colony*, in contrast to an oversimplified model of clonal outbreaks. In fact, these organisms exist in highly diverse populations of related genotypes that reproduce with high mutation rates including point mutations, larger scale insertions and deletions, and gene transfer. Shapiro *et al.* reported ecologically driven differentiation of genes in recently diverged populations of ocean bacteria [5]. While the number of samples in their study was not large enough to observe scale invariance, it demonstrates rapidity with which gene transfer can occur across multiple related strains in response to environmental pressures: "*genomic fragments can sweep through populations in an ecology-specific manner…. with a clear bias toward within-habitat sharing of DNA*" [5].

The theory of quasispecies postulates an evolutionarily optimized mutation rate, which increases in response to stress. This provides for rapid adaptation including, for example, very high rate of drug escape. Moreover, high recombination and horizontal gene transfer rates are known to play an important role in increasing both adaptation rate and genetic diversity. Bacterial populations are known to exhibit high intrinsic mutation and recombination rates, throughout the course of an infection [20] and/or during the acute phase of infection [21]. Increased mutation rates in bacteria are also apparent during selection pressure, either by natural predators [22], or through expression of alternative error-prone mutator genes [23]. Several *Campylobacter* strains are known to show increased mutation rates [24], with instances of *C. jejuni* and *C. coli* hypermutator phenotypes linked to the emergence of ciproxin resistance.

The quasispecies model provides a framework to understand fitness and evolvability of a population [35-37]. Fitness is determined not by the genetic characteristic of a single, isolated, static species or gene, but by the collective distribution of the members of the quasispecies whose clonal expansion reflects their underlying genotypic diversity and their fitness with respect to a particular environment or selection pressure including antibiotics, other medications, and treatments [38,39]. The fallacious model that a particular pathogen associated with an epidemic or outbreak is accurately represented as a single species (e.g., a microbe's name using 16s sequencing) with a clonal identity is not only inadequate, it ignores the underlying genotypic diversity.

Metagenomic techniques make it possible to profile the genetic diversity of a quasi-species within a microbiome. We used RNAseq to study the metatranscriptome of the microbial community of 27 different poultry meal samples. Combining data from all samples provides ~9.5B reads of raw sequence. While metagenomics is often used to profile the community ecology, we used the data here to profile the genotypic diversity of one organism in the community, *Campylobacter*. These reads were aligned to the 218 *Campylobacter* genomes from the Ensembl database and the alignments tallied for each genome at 95% identity. This reference database is larger than is typically used to identity a single species in a metagenomic study, but small enough to illustrate two alternative views of what that identification means.

A classic analysis might treat the 218 *average genomes* in the reference database as independent genotypes, or clonal leaf nodes, on the tree of life. Using this model, if any of these genotypes were present in the community of the poultry meal, it would be possible to set a strict threshold and ignore any evidence of occurrence in the sample below that threshold. For example, selecting a threshold of 500,000 alignments would lead one to conclude that precisely 7 of the 218 genomes in the reference database exist in the sample.

An alternative analysis with a different perspective might treat the 218 genomes in the reference database as approximate, historic, average observations of genotypic *distributions* from 218 ecological niches. Alignments to this reference provide a measure of the probability with which genotypes in the *Campylobacter* quasispecies in the poultry meal are similar to those historic genotypes. Genotypes with several thousands of alignments may reflect minority alleles within the community.

Figure 4 shows a circle plot mapping the log normalized alignments from this metagenomic study of poultry meal to 218 *Campylobacter* reference genomes. A bar chart showing the same data is available in Appendix C. The log normalization highlights the genotypic diversity reflected by both the reference database and the sample. In this view, it is still possible within the circle plot to label specific genotypes with alignment scores above a fixed threshold, but displaying the alignments to the full reference database reveals there is, in fact, a distribution of closely related genotypes within the sample.

As WGS and metagenomics gain acceptance in fields like microbiome medicine, outbreak definition, and food safety traceability, where the demands for accuracy and precision are very high, what really matters is function because it is genes that drive the epidemiology. As shown in Figure 1, hundreds of samples and tens of thousands of individual genes are required to reveal scaling behavior and/or to estimate a power law exponent. To do this accurately requires inclusive and expansive reference databases. If too few references are in the database, inaccurate associations may be made leading to inappropriate attribution of cause and effect. These alignments reveal the genetic diversity of the *Campylobacter* quasispecies within the chicken meal microbiome based on the 218 curated genomes in Ensembl [40].

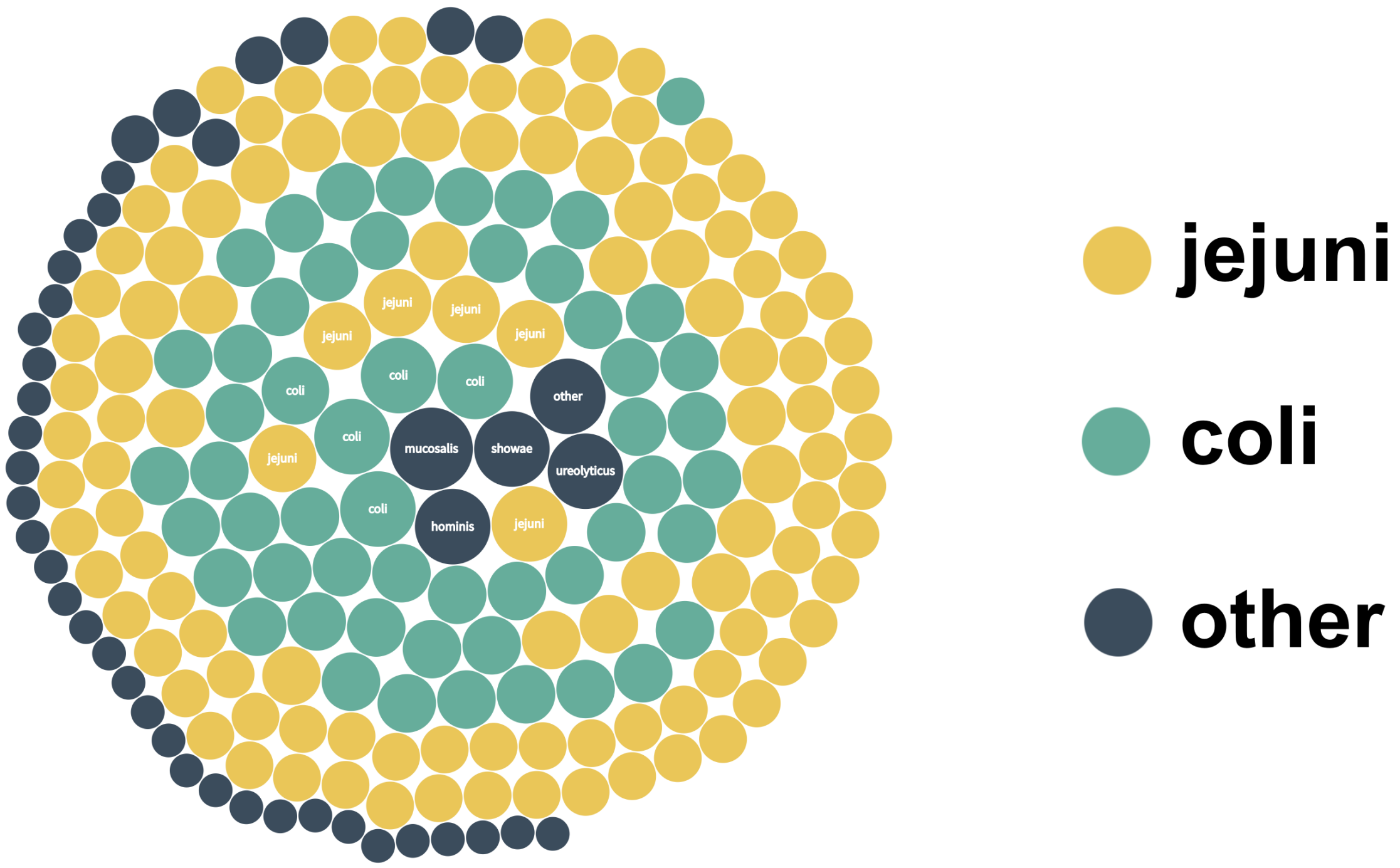

**Figures 4**: A circle plot showing the mapping of all log normalized read alignments from a metagenomic study of poultry meal to 218 *Campylobacter* genomes in the curated Ensembl database. Each circle represents one specific genome in the reference. The area of each circle is the log of the number of sequence reads that matched that genome. The arrangement of the small circles is arbitrary.  The dominant *Campylobacter* species is indicated in the legend.

Another approach to measure diversity in the metagenome is to quantify the allelic variation observed in all reads that align to *Campylobacter*. In the Appendix D we data that suggests the frequency of occurrence of unique SNP variants for all 218 genomes in the reference is approximately a power law with slope near -1 (i.e., *1/f*)

## D. Insular Microbiogeography

The MacArthur-Wilson theory of Island Biogeography holds that there is an underlying power law relationship between species and habitat on a macroscopic level. Our analysis leads us to posit that the power law exponent that MacArthur and Wilson measured for macroscopic life, should also apply to microbes. This is, by definition, scale-free behavior:

*Each host provides a unique environment for community of microorganisms. Microorganisms adapt to their environments.*

*If the distribution of host organisms obeys the MacArthur-Wilson law, so too will the genomic diversity of the microbial communities that colonize them.*

It remains to posit an ecological model connecting the exponent we measured to the MacArthur-Wilson exponent. Our work shows the number of new genes vs. the number of samples or genotypes (Figure 1), whereas MacArthur and Wilson's law relates species or taxa to *geographic area*. Moreover, the data we used comes primarily from sources originating from human patients and, if the human hosts are interpreted as a single niche, there may seem to be no reason to expect scaling behavior or any diversity within the genotypic clades. Contrary to this expectation, we find the data indicates a possibly universal power law relationship as a function of number of isolates. In fact, medical sequencing of pathogens is indicated in the event of human disease, typically in unusual cases. As such, these reference sequences do not specifically represent *steady state populations*, but rather originate from consumption of contaminated food or water, or exposure to other disease vectors. Given a global food supply chain and global travel patterns, the simplest interpretation of the data is that these genomes represent a *near random sampling of foodborne pathogens* globally. Humans are engaged in this random sampling as part of everyday consumption. With random sampling, the number of isolates (or genotypes), $\eta$, in Figure 3, is related to area, A, as a random walk through sample space. Formally, this implies a functional relationship between area and the number of isolates as $A \sim \eta^{1/2}$. The MacArthur-Wilson exponent, $z$, is then one-half the exponent obtained in Figures 1, or:

$$N \propto \eta^{2z} \qquad \text{Eq. 4}$$

For all organisms studied here, the data puts the corresponding MacArthur-Wilson exponent in the range $0.23<z<0.32$, well within the range observed for species-area scaling in the theory of Island Biogeography.

Equation 4 and the data in Figure 1 indicate that as the number of genomes studied increases, the cumulative number of genes observed will continue to increase. The more genomes included in an analysis, the more genes will be discovered [41]. Although the three experiments involved tens of thousands of genomes, none provided evidence for an upper limit to the rate of gene discovery. Such a limit would be manifest as a "knee" or bend on the log-log plots above some maximum of genomes. For example, in Figure 1c, with 15,158 genomes and 1000 jackknife trials, the study spans a scale exceeding four orders of magnitude with no upper scaling limit found. These observations in three different organisms raise the question whether a core genome exists.

For any taxonomic group, at the measured rate of gene discovery, the question becomes "*how many samples would be required before an alternative gene is discovered for every* *gene in any randomly selected genome?"* This is a special case of the "*Coupon Collectors Problem*" [42]. If we assume that non-synonymous mutations are possible for any gene, then the question becomes, after finding $N$ new genes at random from a genome of length $N_o$, at what value of $N$ does the probability of *not finding* an alternative gene for every gene in the original genome become vanishingly small. The derivation is in the Appendix E. The probability of not finding an alternative to *every gene* falls below $1/N_o$ for $N > 2 N_o \ln(N_o)$. For *Campylobacter*, alternatives to all $N_o$ genes in the genome should be found with just a few hundred randomly chosen isolates. For *Salmonella,* observing a variant to each of $N_o$~*4500 genes* requires approximately 2000 isolates.

The existence of scaling over several orders of magnitude in sample diversity of microbial genotypes and genomes has practical implications for microbiology and the way we catalog organisms and investigate outbreaks. Approaches that depend on definition of a core genome, selection of small numbers of SNPs, or of core MLST gene sets, are destined to fail with false negatives. The fact that adding one more genotype to the set can indicate a need to rebuild the taxonomy shows that the tree of life for bacteria is not a tree – it is a network (or graph) of interacting genomes within an environment or geography [28-30,43], with loops that indicate gene transfer between divergent taxa.

As in virology, it is more appropriate to think of bacteria in terms of quasispecies with a power law distribution of genotypes and genes. The tools required for whole genome comparison already exist and must be applied to capture genetic similarity between both nearby and distant genotypes. This is necessary for accurate identification and functional association required by applications in microbiome-induced physiology, outbreak detection, and food regulation. Adopting whole genome comparison may force microbiology to replace the practice of naming and even the concepts of "genus" and "species" with new data models and notations that label populations by their function and the niche(s) they occupy. From a regulatory perspective it is most important to know if an organism has a pathogenic gene that may cause illness (regardless of its taxonomic classification).

There are several desiderate one might consider for a new nomenclature system. These notations should capture the diversity or statistical distributions that describe the frequency with which important function exists within a community. Community level techniques including metagenomics and meta-transcriptomics can be used to predict the probability that specific genes are (or are not) present in a community and the likelihood with which they may emerge as hazards in response to fitness pressures.

## IV. CONCLUSION

The MacArthur-Wilson theory of insular biogeography applies to microbes. Species require niches to survive and thrive. With respect to microbes these niches are not restricted to the physical and chemical properties of insular geographic environments (which themselves evolve a characteristic fractal geometry) but also include other species. It is the interacting web of ecological dependence between species that leads to evolution of scale-free behavior in the diversity of higher-level life forms within isolated ecosystems. These host organisms also provide unique habitats for the communities of microbes that colonize them.

The definition of scale-free behavior implies that the same exponent observed for higher-level organisms in isolate communities also applies at the level of microbes, microbial communities, and the microbial genes (metagenomes) of these hosts. Scale-invariant genetic diversity requires that we approach organism classification with whole genome comparison and an appropriate model of bacterial quasispecies. Evidence for the universality of the MacArthur-Wilson exponent may be found in sequence data from isolates, in assembled genomes from reference databases, and in metagenomic studies. Relating the exponent obtained by enumerating genomes and genes to the exponent obtained from measured land area requires a statistical model of how a microbiome samples area (i.e., how is it fed). A random sampling model suggests the exponent for the scaling of gene diversity is approximately twice the MacArthur-Wilson exponent.

The diversity predicted by a theory of insular microbiology requires an overhaul of our approach to organism classification and regulation. Bacteria form quasispecies related by an ever evolving phylogenetic graph – not by a tree – that describes a complex web of interdependence spanning all phylogenetic ranks down to the level of microbes. In any open ecosystem or open microbiome, the classical concept of core genomes ceases to be meaningful, for any class of bacteria. Classification and regulation must be based on comparison of whole genomes with a view toward identifying pathways and functions with hazardous (or beneficial) potential.

### E. Acknowledgements:

The authors would like to acknowledge contributions from and discussions with Judith Douglas, Nugget Kong, Bob Baker, Peter Markwell, Dylan Storey, Barbara Jones, and Kenneth L. Clarkson.

### F. Competing financial interests:

None of the authors have competing financial interests or other conflicts of interest with respect to this work.

## APPENDIX A: METHODS

All WGS data are publicly available via NCBI [3]. Accession numbers of the genomes used are provided with the supplementary material. The WGS was downloaded, assembled using ABySS (abyss-pe v.1.5.2) [44], and annotated using Prokka (v.1.10) [45] prior to the analysis. These include new WGS data, sequenced by the 100K Pathogen Genome Project (UC Davis, Davis, CA) as described by Ludeke et al. [46], using Illumina paired end 100 methods, publically released on the SRA in the 100K Pathogen Genome Project bioproject (PRJNA186441), and re-assembled for use in this study. To measure the cumulative rate of observation of “new” genes (both known and “putative” or unknown) as a function of number of *Salmonella* isolates sampled, 866 individual *Salmonella* isolates were obtained from the 100K Pathogen Genome Project with *de novo* assembly with random subsets of those isolates selected in separate trials, followed by cumulative gene count determined for each trial. To measure the cumulative rate of observation of “new” *Campylobacter* genes as a function of number of isolates sampled, the raw sequence data from 15,158 individual *Campylobacter* isolates from public sources [18], and assembled and annotated as described for *Salmonella*. The number of genomes represents all available *Campylobacter* WGS data in the SRA, at the time of this publication, for which assembly was successful. The large number was chosen to test the observed scale free behavior over four orders of magnitude in the boot strap as a function of number of genomes.

Identification of core genes to build the phylogeny for *E. coli* was done with the Basic Local Alignment Search Tool (BLAST) with a criterion of 95% identity over 90% of gene length querying 42 closed reference genomes. An allowance was made whereby a core gene was retained if missing in no more than one genome. Alignment of the 3,348 strain sequences for each gene yielded a SNP set that was used to construct a phylogeny. From the phylogeny, 334 genotypes were selected to represent the diversity in the tree.

Whole RNA metagenomes were produced using HiSeq 3000/4000 instruments with one sample per lane resulting in ~350M reads/sample. Genomic distances were determined using the Meier-Kolthoff method [47]. Whole Genome-Genome distance matrices were translated into the Newick tree format using Mega7 with the Neighbor-Joining method. Genome distance matrixes were clustered and visualized using Matlab and R [48]. Genome-genome similarity is defined as 1.0-genome/genome distance.

## APPENDIX B: GENOME-GENOME SIMILARITY FOR E. COLI AND SALMONELLA ENTERICA

Figure 2 of the manuscript reveals similarity for well separated (off-diagonal) *Campylobacter* genome pairs, indicating similarity between divergent branches of the taxonomic graph. The same behavior is observed for *E. coli* and *Salmonella enterica* as evidence by FDA microarray data. Figure B1(a,b) shows whole genome array hybridizations for 1094 *E. coli* genomes from the FDA *E. coli* Identification (ECID) microarray (Panel a) [49], and 600 *Salmonella enterica* profiled on a *S. enterica–E. coli* (SEEC) multispecies microarray (Panel B) [50], The majority of *E. coli* and *Salmonella* profiled by this method were amalgamated from clinical, foodborne, and associated environmental strains with relevance to public health and food safety. Strains from private, academic, and publically available collections were also included. The microarray profiling program was developed by the FDA for track-and-trace molecular epidemiology [50]. It was borne out of an unconventional application towards genomic profiling from its original utility as a gene expression tool. This is primarily because forensic strain-level attribution became possible from the significant intraspecies genome plasticities observed in the resulting profiles [51]. In aggregate, such whole genome analyses revealed that individual species strains could differ on the order of megabases.

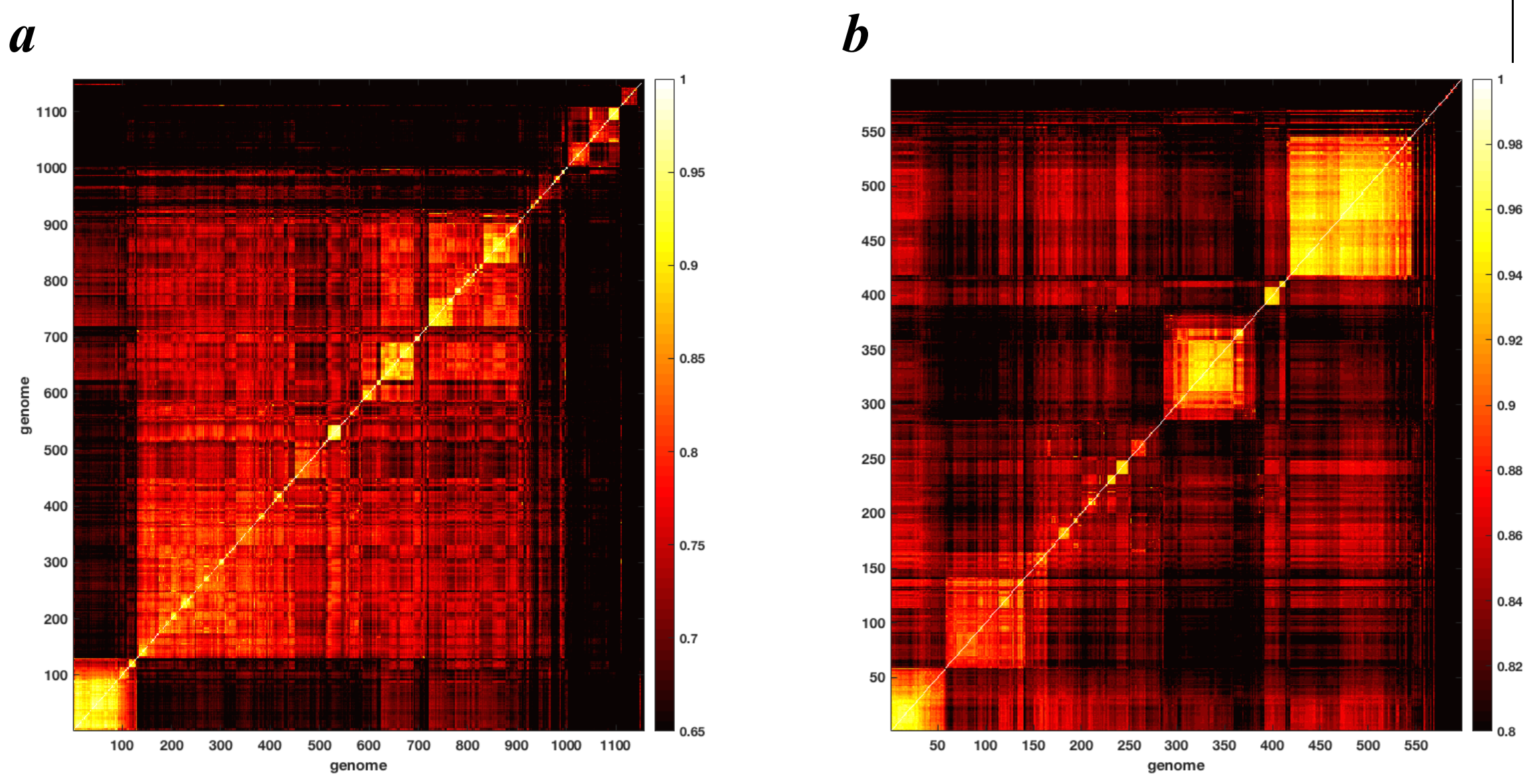

**Figure B1(a,b):** Heat maps of genome-genome similarity (Pearson correlations) of pangenome gene (allele) presence measured by with custom designed Affymetrix microarray platforms. **Panel a**. Shows hybridization intensities for 1094 *E. coli* genomes. **Panel b**. shows hybridization intensities for 600 *Salmonella* enterica genomes.

## APPENDIX C: THE QUASISPECIES MODEL

As discussed in the manuscript, a conventional view (Figure C1a) treats the 218 *average genomes* in the reference database as independent genotypes, or clonal leaf nodes, on a taxonomic graph. Using this model, if any of these genotypes were present in the poultry meal metagenome, it would be possible to set a strict threshold (e.g., the red line at 500,000 alignments) and ignore any evidence of occurrence in the sample below that threshold. The bar heights above threshold in Figure C1a would seem to indicate only a small number of genotypes present in the poultry meal.

Figure 4b expands the y-axis scale to show the same data from a perspective. In this view, read alignments numbering in the hundreds are also meaningful. The 218 genomes in the reference database should be considered approximate, historic, average observations of genotypic *distributions* from 218 *other* ecological niches. No clonal copies of the genomes in the reference are expected in the sample. Alignments to the reference provides a measure of the probability with which genotypes in the *Campylobacter* quasispecies in the poultry meal are *similar* to those historic reference genotypes. Genotypes with several hundred alignments reflect minority alleles within the community.

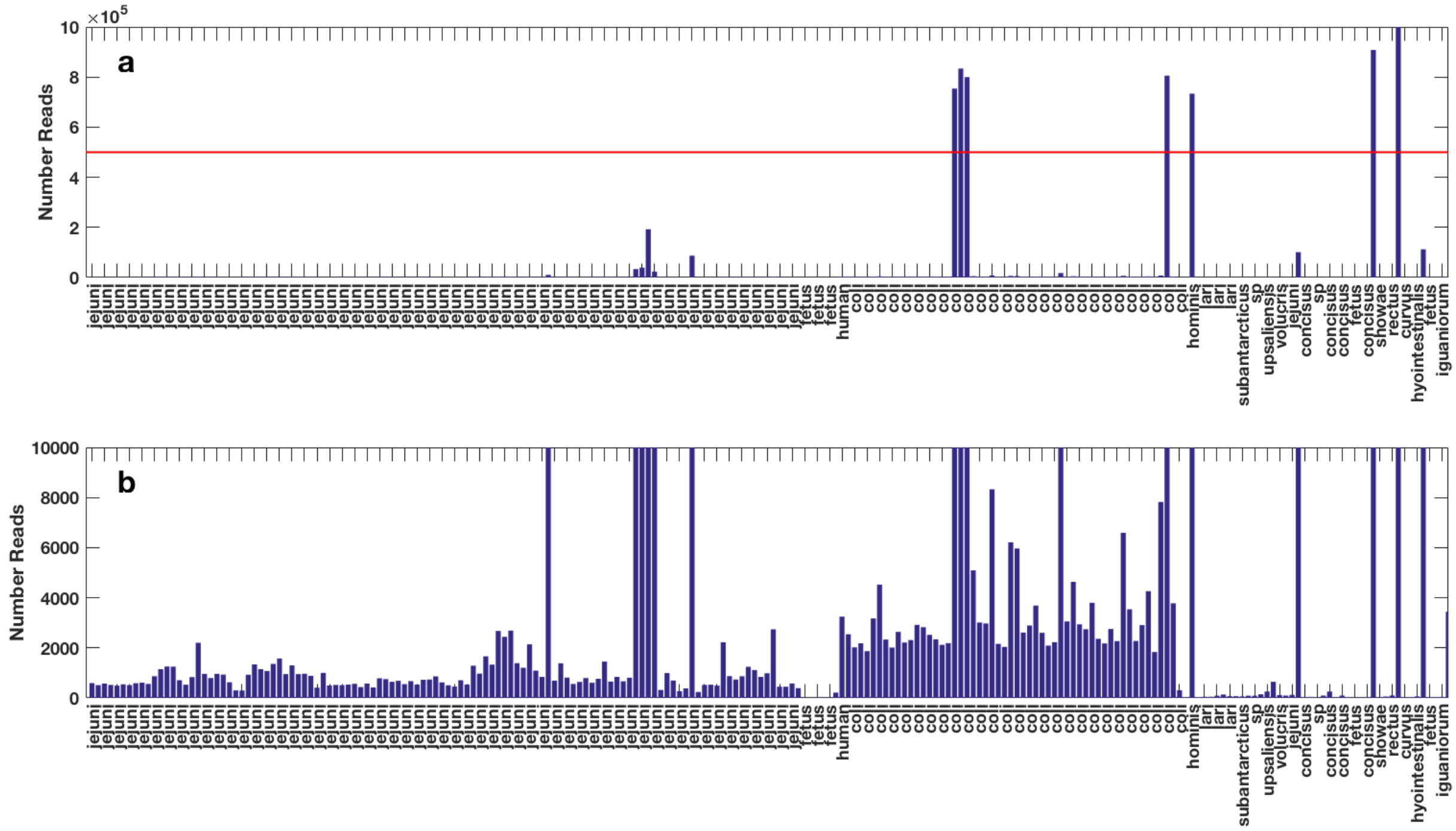

**Figure C1(a,b):**: A mapping of all reads from a metagenomic study of poultry meal to 218 *Campylobacter* genomes based on two different perspectives. Figure C1a sets a threshold to identify a small number of taxa "present in the sample". Figure C1b shows the probability that other genotypes are represented as minority members in the quasispecies distribution.

# APPENDIX D: SNP VARIANT FREQUENCY

Yet another approach to measure diversity is to quantify the allelic variation observed in all reads that align to *Campylobacter*. To do this we enumerate all specific SNP variants and plot in Figure D1 the frequency with which these occur. A particular *Campylobacter* SNP variant (where the nucleotide differs from the reference) that is found only once at a particular location is the most frequent event in the graph. Other variants, where the same substitution is found up to tens of thousands of times at a particular location, are found infrequently. The unique SNP variants were enumerated for all 218 genomes in the reference. The frequency of occurrence for the number of unique SNP variants is approximately a power law with slope near -1.

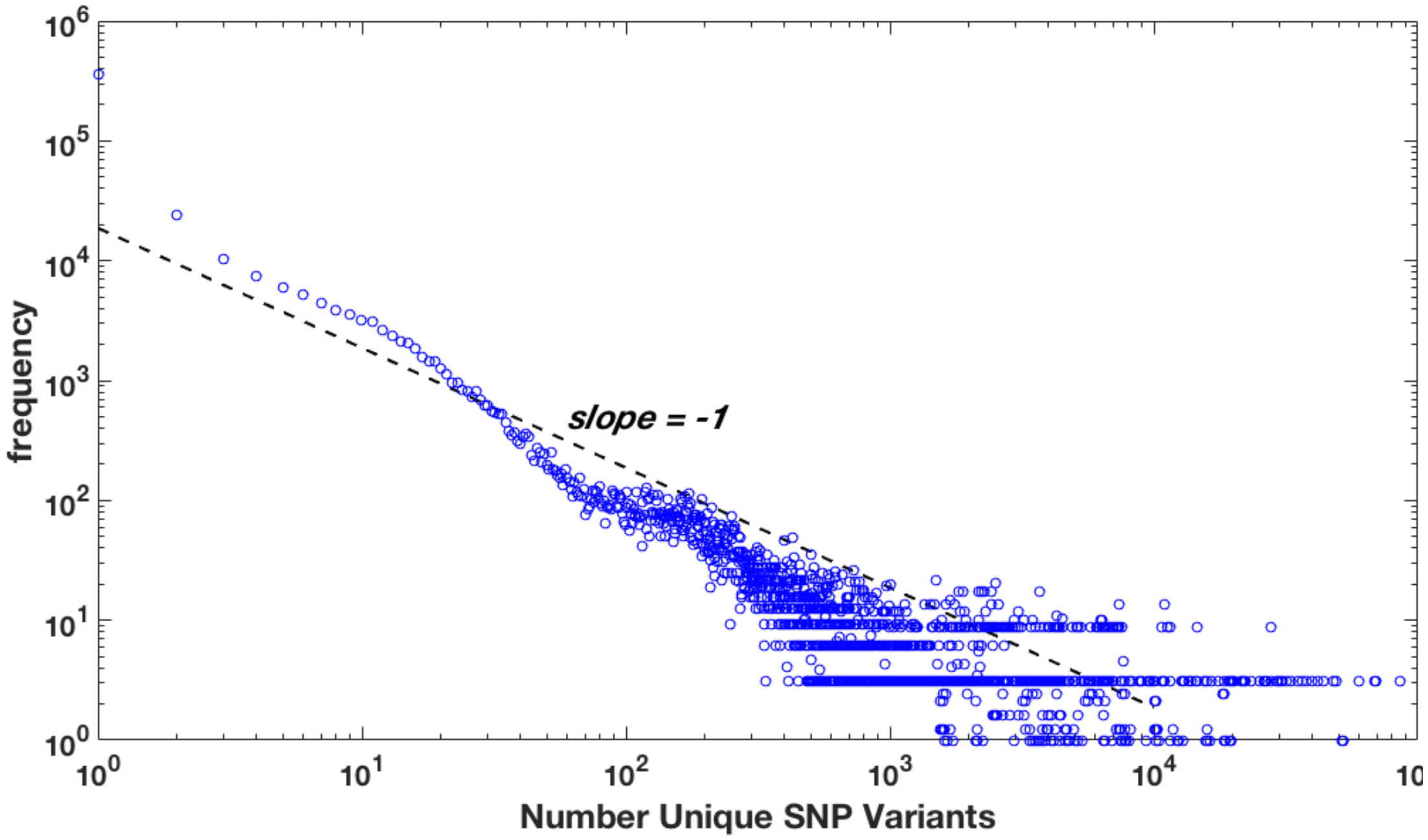

**Figure D1:** The occurrence frequency of unique *Campylobacter* genome SNP variants within a metaRNAseq study of a poultry meal. The function is approximately a power law with slope near -1.

We note that the evidence for scale invariance down to the level of single SNPs does not prove that all nucleotides within a genome are replaced with the same frequency or with the same scale-free behavior. Rather, the highest rate of substitution is typically observed near the 3'-ends of a gene. Substitutions near the 3'-ends are less likely to disrupt function, and more likely to survive (by natural selection).

## APPENDIX E: GENE REPLACEMENT FREQUENCY

The data in manuscript Figure 1a-c reveals that the more genomes we sequence, the more genes will be discovered. This observation begs the question of what a core genome is and how many organisms might be needed to defeat the concept. For any taxonomic group one may ask, at the measured rate of gene discovery, "*how many samples would be required before an alternative gene is discovered for every gene in any randomly selected*

*genome?"*

The answer to this question if given as a special case of the "*Coupon Collectors Problem*" [42]. If we assume that non-synonymous mutations are possible for any gene, then the question becomes: after finding *N* new genes at random from a genome of length $N_o$, at what value of *N* does the probability of *not finding* an alternative gene for every gene in the original genome become vanishingly small.

Let $Z_i^N$ represent the event that the *i*-th gene is not replaced after *N* new genes are found.

$$P(Z_i^N) = (1 - \frac{1}{N_o})^N \leq e^{-N/N_o} \quad \text{Eq. S1}$$

where the inequality follows from the Taylor expansion of $e^{-x}$ for small *x*.

Then for $N > 2\, N_o\, ln(N_o)$

$$P(N > 2\, N_o\, ln(N_o)) = P(\cup_i Z_i^{2\, N_o\, ln(N_o)}) \leq N_o P(Z_i^{2N_o\, ln(N_o)}) \leq \frac{1}{N_o} \quad \text{Eq. S2}$$

For *Campylobacter*, with $N_o$~1500 genes, this requires finding N~ 22000 new genes. From Figure 1 this requires will take place after randomly sampling just a few hundred isolates. For *Salmonella,* a variant to each of *$N_o$~4500 genes* will be found after discovering ~76,000 genes which requires approximately 2000 isolates. *The implication is that a unique core genome does not exist.*